\documentclass[aps,superscriptaddress,showpacs,showkeys,nofootinbib,twocolumn,floatfix]{revtex4}
\usepackage{graphicx,epsfig,wrapfig,color}
\usepackage{color}

\definecolor{darkgreen}{rgb}{0,0.65,0}

\newcommand{\be}{\begin{equation}}
\newcommand{\ee}{\end{equation}}
\newcommand{\ba}{\begin{eqnarray}}
\newcommand{\ea}{\end{eqnarray}}
\newcommand{\di}{\!{\rm d}}
\newcommand{\la}{\langle}
\newcommand{\ra}{\rangle}
\begin{document}
\newcommand*{\Yale}{Department of Physics Yale University, 
   New Haven, CT 06511-8499, U.S.A.}\affiliation{\Yale}
\newcommand*{\UConn}{
  Department of Physics, University of Connecticut,
  Storrs, CT 06269, U.S.A.}\affiliation{\UConn}
\title{\boldmath
  Radial excitations of $Q$-balls, and their $D$-term}
\author{Manuel Mai}\affiliation{\Yale}
\author{Peter Schweitzer}\affiliation{\UConn}
\date{June 2012}
\begin{abstract}
We study the structure of the energy-momentum tensor of radial 
excitations of $Q$-balls in scalar field theories with U(1) symmetry. 
The obtained numerical results for the $1\le N \le 23$ excitations 
allow us to study in detail patterns how the solutions behave with $N$. 
We show that although the fields $\phi(r)$ and energy-momentum tensor 
densities exhibit a remarkable degree of complexity, the properties
of the solutions scale with $N$ with great regularity.
This is to best of our knowledge the first study of the $D$-term $d_1$ 
for excited states, and we demonstrate that it is negative --- 
in agreement with results from literature on the $d_1$ of ground
state particles. 
\end{abstract}
%
%
%
\maketitle
\section{Introduction}
\label{Sec-1:introduction}

The energy momentum tensor $T_{\mu\nu}$ (EMT) is a central 
quantity in the field theoretical description of particles.
Its matrix elements \cite{Pagels} give the mass \cite{Ji:1994av}, 
the spin \cite{Ji:1996ek}, and the constant $d_1$ of a particle
\cite{Polyakov:1999gs} to which we shall loosely refer as the 
$D$-term. Though not known experimentally, $d_1$ is 
a particle property as fundamental as mass, spin, electric 
charge or magnetic moment. Its physical meaning is that it gives 
unique insights into the distribution of internal (in hadrons:
strong) forces \cite{Polyakov:2002yz}.

EMT form factors found little practical applications
\cite{Donoghue:1991qv}, until it became clear  that they can 
be accessed by means of generalized parton distribution functions 
\cite{Muller:1998fv,Ji:1998pc} in hard exclusive 
reactions such as deeply virtual Compton scattering
\cite{Adloff:2001cn,Airapetian:2001yk,Stepanyan:2001sm,Munoz-Camacho:2006hx}.
Since that the EMT form factors were investigated in theoretical 
frameworks including chiral perturbation theory, lattice QCD, 
or effective chiral field theories, see 
\cite{Polyakov:2002yz,Polyakov:1999gs} and \cite{Mathur:1999uf,Kubis:1999db,
Petrov:1998kf,Goeke:2007fp,Cebulla:2007ei,Megias:2004uj,Liuti:2005gi}.

Remarkably, in all theoretical studies $d_1$ of pions, nucleons, nuclei
was found negative. A possible explanation of this observation provide
chiral soliton models \cite{Goeke:2007fp,Cebulla:2007ei}, which 
describe the nucleon in the limit of a large number of colors 
$N_c$ in QCD \cite{Witten:1979kh}.
In these models the negative sign of $d_1$ emerges as a natural
consequence of the stability of the nucleon 
\cite{Goeke:2007fp,Cebulla:2007ei}.

To shed some light on the question whether $d_1<0$ is a 
general and model-independent feature, in Ref.~\cite{Mai-new} 
the EMT of $Q$-balls was studied.
These non-topological solitons appear in theories with 
global symmetries, and it is the appearance of the associated 
conserved charge(s) which plays a crucial role for their existence 
\cite{Friedberg:1976me,Coleman:1985ki,Safian:1987pr}.

$Q$-balls have numerous applications in astrophysics, cosmology, 
and particle physics 
\cite{Cohen:1986ct,Alford:1987vs,Lee:1991ax,Kusenko:1997ad,Multamaki:1999an,Kasuya:1999wu,Graham:2001hr,Volkov:2002aj,Clougherty:2005qg,Clark:2005zc,Gleiser:2005iq,Schmid:2007dm,Gani:2007bx,Verbin:2007fa,Sakai:2007ft,Bowcock:2008dn,Tsumagari:2008bv,Arodz:2008jk,Gabadadze:2008sq,Campanelli:2009su}.
They provide an extremely fruitful framework for the purpose of 
clarifying the relation $d_1$ and stability arguments.
In \cite{Mai-new} an extensive study of the EMT structure 
of ground state solutions was presented. In all cases
$d_1<0$ was found, and a rigorous proof was formulated that the
$D$-term of $Q$-balls must be negative.
Moreover, it was shown that stability is a sufficient but not
necessary condition for $d_1$ to be negative, because some 
ground state solutions describe absolutely stable, others meta-stable 
or unstable $Q$-balls, depending on the parameters. The general 
proof applies to all cases and always $d_1<0$ \cite{Mai-new}.

This work is dedicated to the study of the EMT of radial
excitations of $Q$-balls in scalar field theories with U(1) symmetry. 
To best of our knowledge, this is the first study of the $D$-term 
going beyond the description of a ground state. 
Radial excitations of $Q$-balls were studied previously
in \cite{Volkov:2002aj}, where the ground state and the first 
two excited states $N=1, \,2$ were found for a fixed value of 
the charge $Q$. In this work, we will work with a fixed value 
of the angular velocity $\omega$ in the U(1)-space, and study 
the first $1 \le N \le 23$ excitations. 
With $N=0$ denoting ground states, the family of $Q$-ball solutions 
can hence be classified by specifying $(Q,N)$ as done in 
\cite{Volkov:2002aj}, or by specifying $(\omega,N)$ 
as chosen in this work.

Our numerical results reach high in the spectrum of radial
excitations and give fascinating and detailed insights in the 
properties of excited $Q$-balls. In particular, we will see
that also excited states have a negative $d_1$.
The present work extends and completes our study of the EMT 
structure of ground state $Q$-balls. 
It is important to remark that we make no effort to describe
the full spectrum of $Q$-balls which would include also
vibrational or other excitations \cite{Coleman:1985ki}, and 
we will not consider quantum corrections \cite{Graham:2001hr}.

The lay-out of this work is as follows. 
In Sec.~\ref{Sec-2:radial-excitations} we will briefly introduce 
the framework, and review how radial excitations of $Q$-balls emerge 
\cite{Volkov:2002aj}.
In Sec.~\ref{Sec-3:densities} we will present the solutions 
for the ground state and radial excitations $1 \le N \le 23$ 
which we were able to find with our numerical method,
and discuss the charge density and the EMT densities.
In Sec \ref{Sec-4:bulk-properties} we will discuss global properties 
like charge, mass, mean square radii, and the $D$-term 
and investigate patterns how these properties scale
with $N$. Remarkably, among the studied quantities 
$d_1$ varies most strongly with $N$.
Finally, in Sec.~\ref{Sec-V:stability-and-d1} we will focus 
on the issue of stability and the sign of the $D$-term. 
The conclusions will be presented in Sec.~\ref{Sec-6:conclusions},
and some technical questions addressed in Appendices.


\section{\boldmath  $Q$-balls and radial excitations}
\label{Sec-2:radial-excitations}

In this Section we briefly review the theory of $Q$-balls, and
introduce the indispensable formulae on the EMT. We use 
throughout the notation of \cite{Mai-new}, and refer to it
for more details. 
We study the relativistic field theory of a complex scalar field 
$\Phi(x)$ with global U(1) symmetry 
\ba\label{Eq:Lagrangian}
   {\cal L}&=&\frac12\,(\partial_\mu\Phi^\ast)(\partial^\mu\Phi) - V\,,
\ea
where, for suitable potentials $V$ \cite{Coleman:1985ki}, 
$Q$-balls emerge as finite energy solutions of the type 
$\Phi(t,\vec{x}) = \exp(i\omega t)\,\phi(r)$ with $r=|\vec{x}\,|$ 
and $\phi(r)$ satisfying the equation of motion
\ba\label{Eq:eom}
&&   \phi^{\prime\prime}(r)+\frac2r\;\phi^\prime(r)+\omega^2\phi
     - V^\prime(\phi) = 0\:,\\
&&  \phi(0) \equiv \phi_0 \,,   \;\;
    \phi^\prime(0) = 0    \,,   \;\;
    \phi(r)\to 0\;\;\mbox{for}\;\;r\to\infty\;. \nonumber
\ea
We will use the potential $V(\phi)=A\,\phi^2-B\,\phi^4+C\,\phi^6$ 
with $A=1.1$, $B=2.0$, $C=1.0$ \cite{Volkov:2002aj,Mai-new}, 
and set $\omega^2=1.37$ which is among the ideal values for our purposes, 
see App.~\ref{App:technical-details}.

To demonstrate the existence of ground (and excited) $Q$-ball states, 
one can identify $r\to t$ and $\phi(r)\to x(t)$ \cite{Coleman:1985ki}, 
and interpret (\ref{Eq:eom}) as the Newtonian equation for a unit 
mass particle moving under the influence of the friction 
$F_{\rm fric}=-\frac2t\,\mbox{\it\. x}(t)$ in an effective potential 
$U_{\rm eff}=\frac12\omega^2\,x^2-V$, 
\ba
\label{Eq:Newtonian-eom}
\mbox{\it\"x}(t)=F_{\rm fric}-\nabla U_{\rm eff}(x) \,.
\ea
A ground state solution corresponds to the situation that 
the particle starts at $t=0$ from rest at $x_0\to\phi_0$, 
and its motion terminates in the origin $x=0$ after infinite time.

\begin{figure}[t!]
\includegraphics[width=7cm]{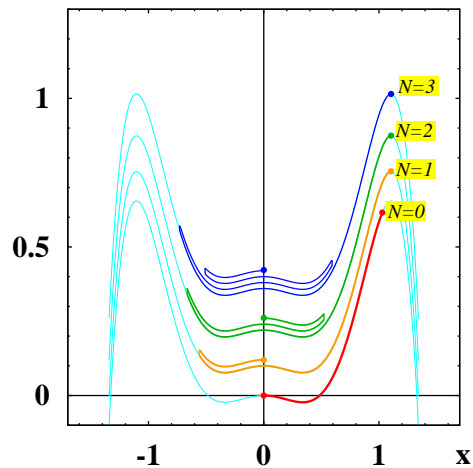}

\vspace{-3mm}

\caption{\label{Fig-01:trajectories-of-radial-excitations}
The effective potential $U_{\rm eff}(x)=\frac12\omega^2x^2-V(x)$ 
as used in this work vs.\ $x$ (thin line). The particle 
trajectories are indicated for $N=0,\, 1,\, 2,\, 3$ (solid lines).
For better visibility for $N=(1,\, 2,\, 3)$ the potentials 
are displaced by $(0.1,\,0.22, 0.36)$ as compared to $N=0$,
and the particle trajectories are displaced by $0.02$
after each turn.} 
\end{figure}

In this picture radial excitations correspond to the situation
when the particle is given more potential energy such that it 
overshoots the point $x=0$, moves ``up-hill'' in the effective 
potential till it reaches a point of return, and finally comes 
to rest at the origin. In principle, the starting points can be 
chosen such that the particle will overshoot the origin 1, 2, 3,
$\dots\;$, $N$ times, see
Fig.~\ref{Fig-01:trajectories-of-radial-excitations}.
This means the corresponding solution $\phi(r)$ has $N$ nodes 
at finite $r$, and we refer to it as the $N^{\rm th}$ 
radial excitation. The ground state correspond to $N=0$.

This picture helps to anticipate several features of the 
excitations. As $N$ increases, the particle has to travel 
longer paths, and do more work against the friction.
Thus we have to release it ``close'' to the maximum 
of $U_{\rm eff}$ where the effective potential is nearly flat, 
see Fig.~\ref{Fig-01:trajectories-of-radial-excitations}.
The particle has to ``wait'' there for a sufficiently long time 
before ``sliding'' down the potential, 
such the time-dependent friction is adequately decreased
to allow the particle to complete its trajectory.

Therefore, as $N$ increases, $\phi_0$ approaches the position of 
the maximum of $U_{\rm eff}$, see App.~\ref{App:technical-details},
and $\phi(r)\simeq\phi_0$ remains basically constant over increasingly 
extended plateaus ``to wait for the frictional force'' to diminish. 
The small-$r$ behavior which follows from (\ref{Eq:eom}) is
\cite{Mai-new}
\be
\label{Eq:asymp-small}
    \phi(r) = 
    \phi_0 - \frac{U^\prime_{\rm eff}(\phi_0)}{6}\;r^2 + 
    \frac{U^\prime_{\rm eff}(\phi_0)\,U^{\prime\prime}_{\rm eff}(\phi_0)}{120}
    \,r^4
    + {\cal O}(r^6)  \;.\;\;\;
\ee
In this Taylor expansion only even powers of $r$ occur, and we 
checked that the coefficients $c_k$ for $k=6,\,8,\,10,\,12$
are also proportional to $U^\prime_{\rm eff}(\phi_0)$ though the
expressions become lengthy. This explains why $\phi(r)$ exhibits
a plateau.
After the plateau we expect $\phi(r)$ to ``oscillate'' $N$-times,
before it vanishes at asymptotically large $r$ according to \cite{Mai-new}
\be\label{Eq:asymp-large}
    \phi(r) \to \frac{c_\infty}{r}\;
    \exp\biggl(-r\sqrt{\omega_{\rm max}^2-\omega^2}\biggr)\,.
\ee
With our numerical method described in 
App.~\ref{App:technical-details} we were able to 
find solutions for the first $N=23$ excited states.

In the following we will discuss the charge density $\rho_{\rm ch}$,
and the EMT densities, namely energy density, $T_{00}(r)$, 
pressure and shear force distributions, $p(r)$ and $s(r)$, 
which are given by \cite{Mai-new} 
\ba
   \rho_{\rm ch}(r)&=&\omega\;\phi(r)^2\,,
   \label{Eq:charge-density}\\
   T_{00}(r)&=&\frac12\,\omega^2\phi(r)^2+\frac12\,\phi^\prime(r)^2+V(\phi)\;,
   \label{Eq:energy-density}\\
   s(r) &=& \phi^\prime(r)^2 \,,
   \label{Eq:shear-density}\\
   p(r) &=& \frac12\,\omega^2\phi(r)^2-\,\frac16\,\phi^\prime(r)^2 -V(\phi)\;.
   \label{Eq:pressure-density}
\ea
We also define the conserved charge $Q=\int\di^3x\;\rho_{\rm ch}(r)$
due to the U(1)-symmetry of the theory (\ref{Eq:Lagrangian}), the 
mass $M=\int\di^3x\;T_{00}(r)$, and the constant $d_1$ which can be 
expressed equivalently in terms of $s(r)$ and $p(r)$ as follows
\be\label{Eq:def-d1}
    d_1 = -\,\frac{1}{3}\,M\int_0^\infty\di^3x\;r^2s(r)
        = \frac{5}{4}\,M\int_0^\infty\di^3x\;r^2p(r)\;.
\ee
The large-$r$ asymptotics (\ref{Eq:asymp-large})
ensures that the integrals defining $Q$, $M$, $d_1$ 
are well-defined.

\begin{figure*}[t!]
\includegraphics[width=16cm]{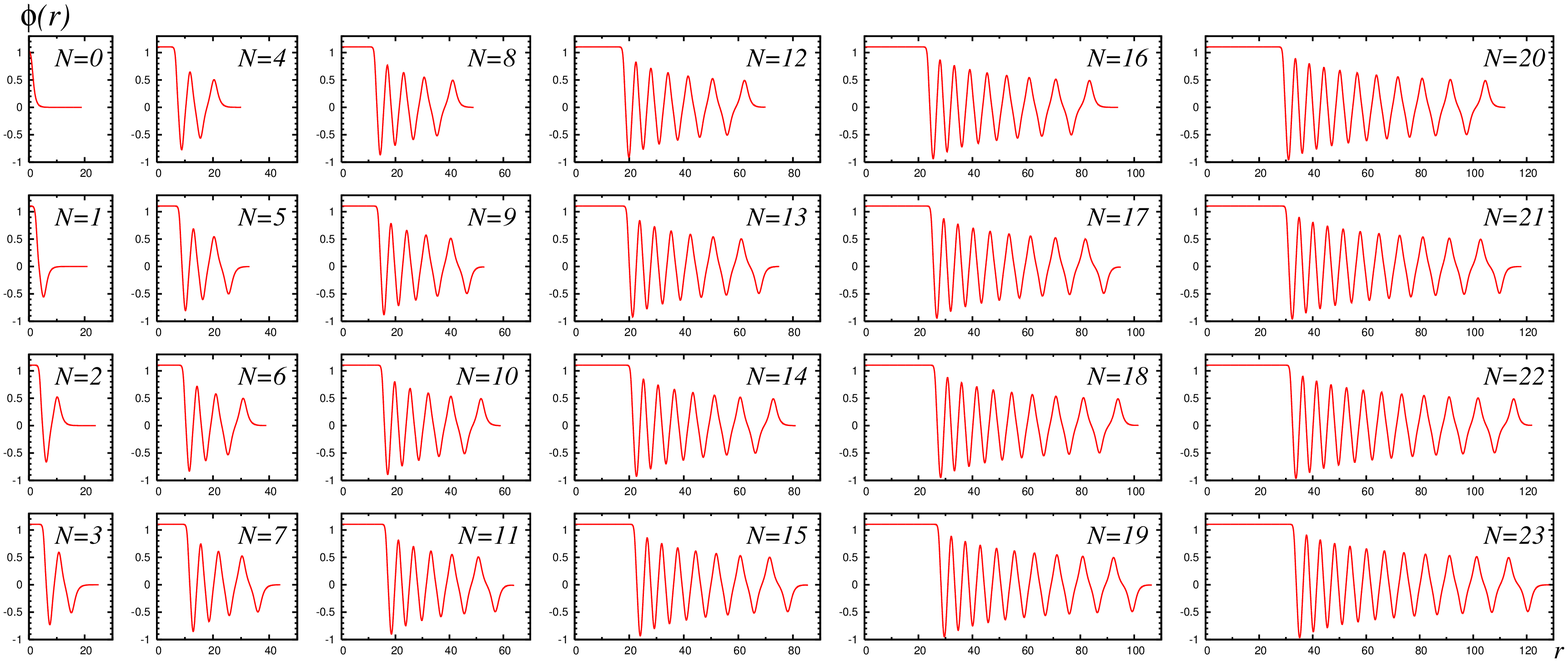}
\vspace{-3mm}

\caption{\label{Fig-02:phi}
   The fields $\phi(r)$ as functions of $r$ for $0\le N \le 23$.}

\vspace{2.5mm}

\includegraphics[width=16cm]{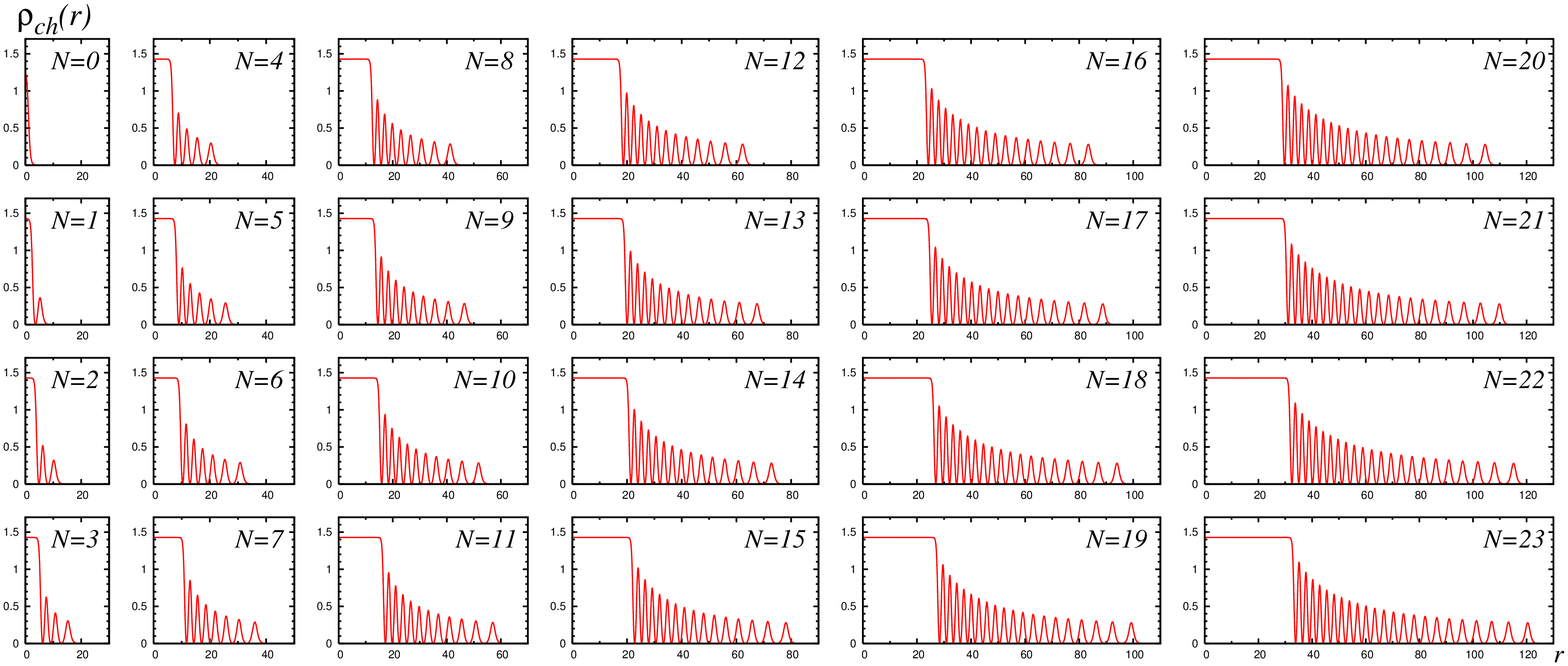}
\vspace{-3mm}

\caption{\label{Fig-03:rho}
   The charge distributions $\rho_{\rm ch}(r)$
   as functions of $r$ for $0\le N \le 23$.}

\end{figure*}

\begin{figure*}[b!]
\includegraphics[width=17cm]{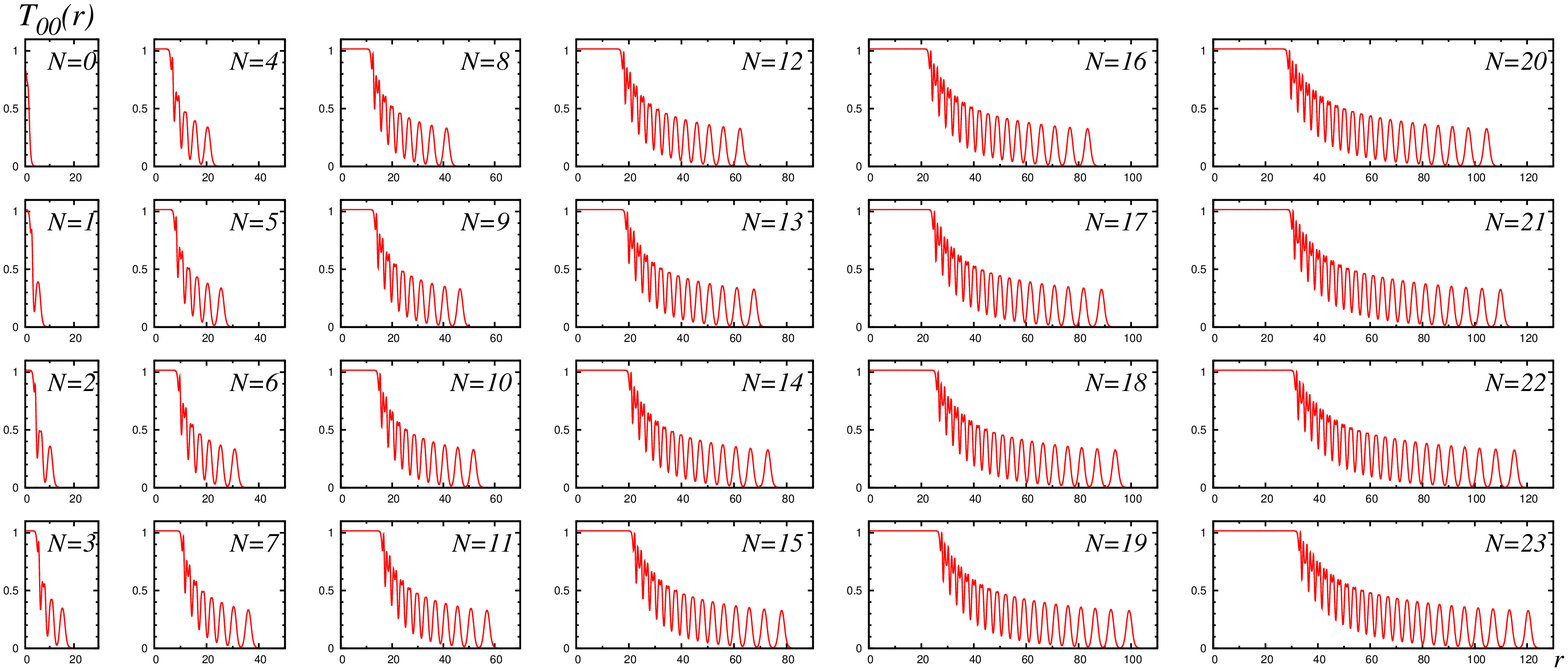}
\vspace{-3mm}

\caption{\label{Fig-04:T00}
  The energy densities $T_{00}(r)$ as functions of $r$ for $0\le N \le 23$.}

\vspace{3mm}

\includegraphics[width=17cm]{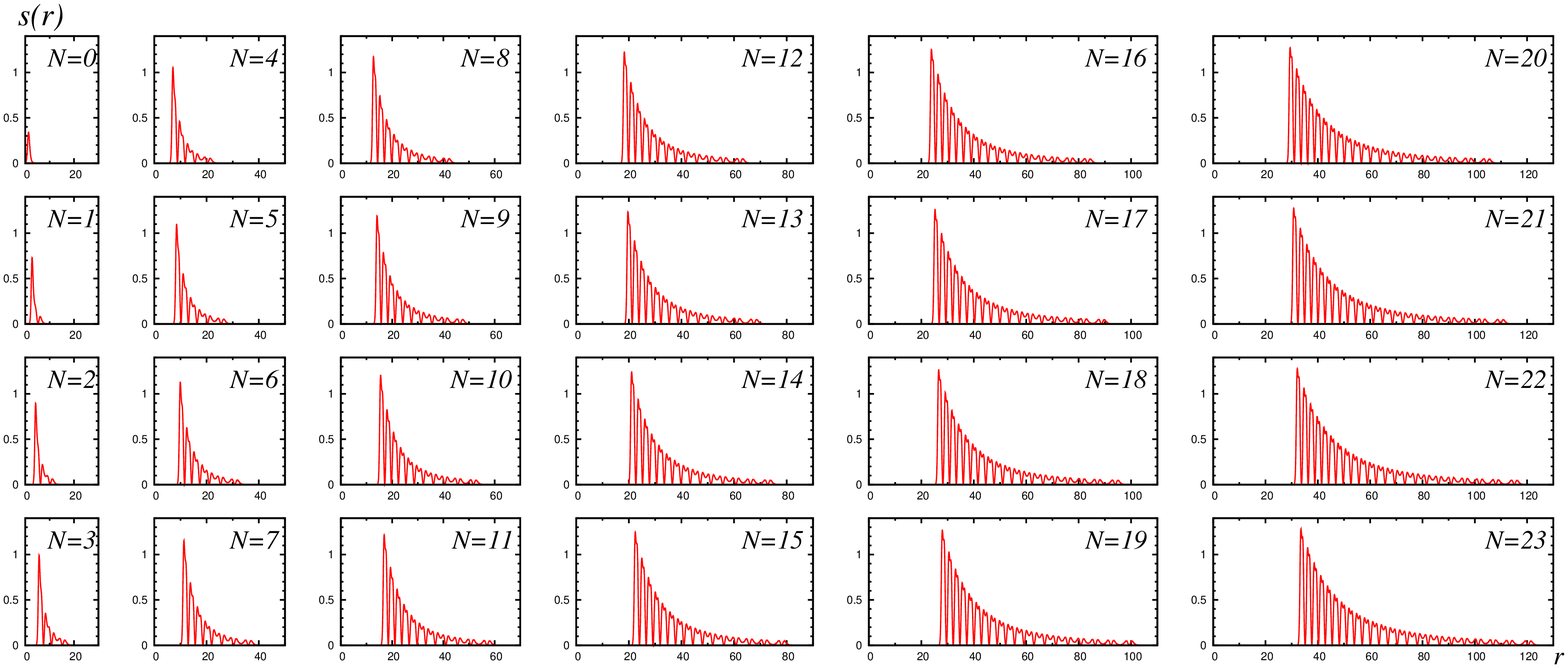}
\vspace{-3mm}

\caption{\label{Fig-05:shear}
  The shear force distributions $s(r)$ as functions of $r$ for $0\le N \le 23$.}

\vspace{3mm}

\includegraphics[width=17cm]{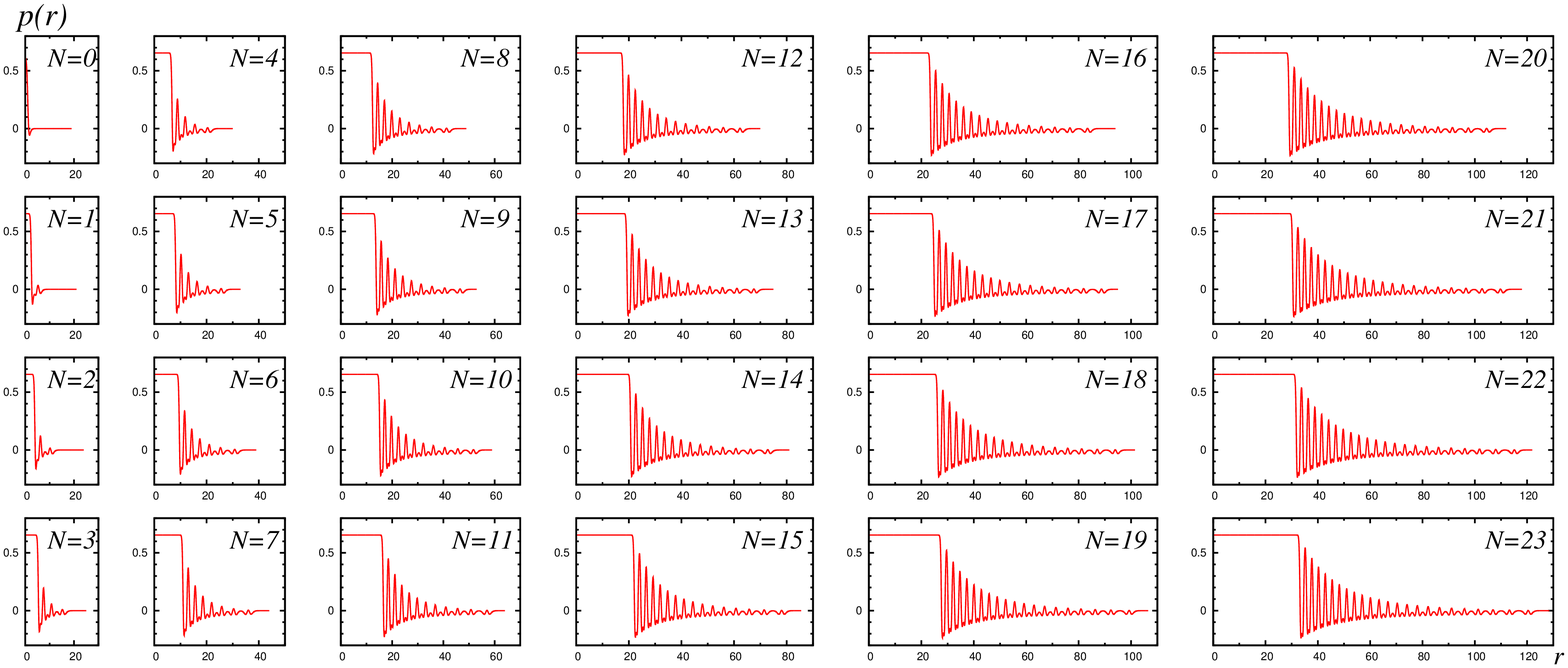}
\vspace{-3mm}

\caption{\label{Fig-06:pressure}
  The pressure distributions $p(r)$ as functions of $r$ for $0\le N \le 23$. }

\end{figure*}

\section{Results for the densities}
\label{Sec-3:densities}

Fig~\ref{Fig-02:phi} shows the results for the radial fields $\phi(r)$, 
the ground state $N=0$ and radial excitations $1\le N\le 23$. 
The results confirm the features we predicted in 
Sec.~\ref{Sec-2:radial-excitations}. 
For $N>0$ the initial values $\phi_0$ are numerically within 
$10^{-6}$ close to each other. For $N\gtrsim 2$ the solutions 
show plateaus with $\phi(r)\simeq\phi_0$, followed by regions 
of ``oscillatory behavior'' with $N$ zeros, before the exponential 
decays set in according to (\ref{Eq:asymp-large}). 
For $N\gtrsim 4$ the sizes of the plateau regions and oscillatory 
regions are roughly in a constant $1\,:\,3$ ratio.

The $N$ zeros of the solutions $\phi(r)$ imply a strict shell 
structure for the charge distributions $\rho_{\rm ch}(r)$ which 
is shown in Fig.~\ref{Fig-03:rho}. The $N^{\rm th}$ excited state 
consists of an inner region of nearly constant charge density
for $N\gtrsim 4$, followed by an outer region with $N$ shells. 

Also the energy densities $T_{00}(r)$ in Fig.~\ref{Fig-04:T00} 
exhibit characteristic shell structures. Although they never 
vanish at finite $r$, the $T_{00}(r)$ show noticeable minima
numerically very close to the zeros of $\rho_{\rm ch}(r)$.
This can be understood in the particle motion picture as follows. 
We have $T_{00}^\prime(r)=
\frac{\partial\;}{\partial t}E_{\rm kin}$
for $r\in\{R_i \,|\, \phi(R_i)=0, \; 1\le i\le N\}$,
i.e.\ the positions $R_i$, where the fields and hence also 
charge distributions vanish, correspond in time to the transits of 
the particle through the origin, and $T_{00}^\prime(R_i)$ correspond 
to time-derivatives of the kinetic energies at those times. 
In the absence of frictional forces $E_{\rm kin}$ would be exactly 
extremal at the origin. Because of friction the extrema of 
$E_{\rm kin}$ are somewhat shifted, but those shifts decrease with 
time ($\leftrightarrow$ distance) because $F_{\rm fric}\propto\frac1t$.

For $N\gtrsim 2$ the energy densities show ``spikes'' at the edge 
of the inner bulk region. For $N\gtrsim 3$ also the subsequent 
inner shells exhibit characteristic ``double-spike'' structures.
The reason for that is the contribution of the surface energy
\cite{Mai-new}. The concepts of surface tension and surface
energy are well defined for $\omega\to\omega_{\rm min}$ 
\cite{Coleman:1985ki}, but the associated features are 
noticeable also away from this limit \cite{Mai-new}.
If the inner region and the $N$ shells had sharp edges, 
$s(r)$ would consist of $(2N+1)$ $\delta$-functions marking the 
positions of the respective surfaces. For our parameters the
system is diffuse, but the ``smeared out $\delta$-functions'' 
in $s(r)$ can be seen in Fig.~\ref{Fig-05:shear} though the 
``gaps'' between the first shells cannot be clearly resolved.

Also this can be understood in the particle picture, where
$s(r) \to 2E_{\rm kin}(t)$. The zeros of $s(r)$ coincide with the 
turning points in Fig.~\ref{Fig-01:trajectories-of-radial-excitations}.
The maxima of $s(r)$ occur at the positions where the particle is fastest,
which is close to the origin of the particle coordinate\footnote{
  To recall, the origin in the particle coordinate $x(t)$ corresponds 
  to the zeros of $\phi(r)$. The latter are also the zeros of the charge 
  distribution and close to the minima of $T_{00}(r)$, see above, which 
  emphasizes that all quantities reflect the same shell structure.} 
in Fig.~\ref{Fig-01:trajectories-of-radial-excitations}.
The characteristic double peaks emerge 
because the particle is slowed down at the origin by the 
buckle in $U_{\rm eff}$. At earlier times (inner region) the friction 
$F_{\rm fric}\propto \frac{1}{t}$ is noticeable making the double peaks 
less symmetric and hard to resolve, see Fig.~\ref{Fig-05:shear}.
At later times (outer region) the friction is diminished, and 
the double peaks are nearly symmetric. 

Fig.~\ref{Fig-06:pressure} show that the pressure distribution 
of the $N^{\rm th}$ excitation changes the sign $(2N+1)$ times. 
Although with increasing $N$ the structures are more and more 
complex, the results are numerically stable and satisfy the 
stringent tests discussed in App.~\ref{Sec:test-numerics}.
In particular, in all cases the stability condition is satisfied
within numerical accuracy, as we will discuss in detail in 
Sec.~\ref{Sec-V:stability-and-d1}.

\begin{figure}[h!]
\includegraphics[width=7cm]{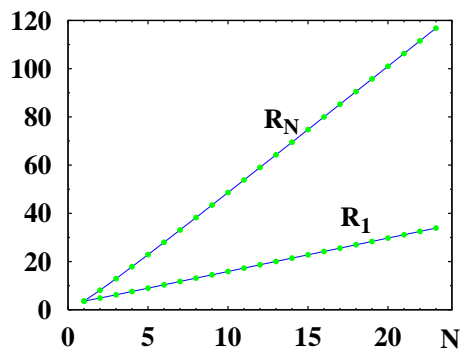}

\vspace{-3mm}

\caption{\label{Fig-07:Rmin-Rmax}
The positions of the first ($R_1$) and last ($R_N$) zero of the
$N^{\rm th}$ radial excitation as function of $N$ for $1\le N\le 23$.
The discrete data sets are connected by lines to guide the eye.}
\end{figure}

Figs.~\ref{Fig-02:phi}--\ref{Fig-06:pressure} demonstrate that
with increasing $N$ the system becomes larger and exhibits an 
increasing degree of complexity. 
In spite of the complexity, however, the size of the system 
grows with remarkable regularity, as is shown in 
Fig.~\ref{Fig-07:Rmin-Rmax}.
This Figure displays for the excitations $1 \le N \le 23$ the respectively
first ($R_1$) and last ($R_N$) zero of the solutions $\phi(r)$.
For $N=1$ the two radii coincide. We observe that the $R_1$ and 
$R_N$ increase linearly with the order of the excitation.

\newpage
\section{Global properties}
\label{Sec-4:bulk-properties}

Above we made three important observations which will allow 
us to make predictions for the $N$-behavior of the global 
(integrated) properties of $Q$-balls, namely
\begin{enumerate}
\item[\sl (i)]  the system exhibits a shell structure,
\item[\sl (ii)]  the size of the system grows linearly with $N$, 
\item[\sl (iii)]  $\rho_{\rm ch}(r)$ and $T_{00}(r)$ inside the $Q$-balls 
                   are effectively constant independently of $N$.
\end{enumerate}
The shell structure of point (i) is evident from 
Figs.~\ref{Fig-02:phi}--\ref{Fig-06:pressure}.
The linear growth of point (ii) is apparent from 
Fig.~\ref{Fig-07:Rmin-Rmax}. Point (iii) however 
requires some explanation.
Strictly speaking the fields $\phi(r)$ and consequently $\rho_{\rm ch}(r)$
and $T_{00}(r)$ are constant only in the inner region, i.e.\ in about 
1/4 of the size of excited $Q$-balls. However, when integrating we 
effectively ``average'' over the oscillatory behavior of these densities 
in the outer region. Therefore, when speaking about global
(integrated) properties
we may think in terms of effectively constant densities inside 
excited $Q$-balls which motivates assumption (iii).

On the basis of these observations we expect the following $N$-behavior
of the charge $Q$, mass $M$, constant $d_1$, the surface tension $\gamma$,
surface energy $E_{\rm surf}$, and the mean square radii 
$\la r_Q^2\ra$, $\la r_E^2\ra$, $\la r_s^2\ra$
of respectively the charge, energy, and shear force distributions:
\ba
                 Q &\propto& N^3\label{Eq:N-behave-Q} \\
                 M &\propto& N^3\label{Eq:N-behave-M} \\
                d_1&\propto& N^8\label{Eq:N-behave-d1} \\
            \gamma &\propto& N  \label{Eq:N-behave-gamma}\\
      E_{\rm surf} &\propto& N^3\label{Eq:N-behave-Esurf}\\
\la r_i^2\ra^{1/2} &\propto& N  \label{Eq:N-behave-ri2}
       \, , \;\;\; i = Q, \;E,\; s\,.
\ea
The surface energy is given by $E_{\rm surf}=\int\di^3r\,s(r)$,
while $\la r_Q^2\ra=\int\di^3r\,r^2\rho_{\rm ch}(r)/Q$ and
$\la r_E^2\ra$ is defined analogously. 
Finally, the mean square radius of the shear forces is 
$\la r_s^2\ra=\int_0^\infty\di r\,r^2s(r)/\gamma$ where
$\gamma=\int_0^\infty\di r\,s(r)$ denotes the surface tension.
Surface energy and surface tension are well motivated notions
in the limit $\omega\to\omega_{\rm min}$ \cite{Coleman:1985ki} 
in which $Q$-balls behave like liquid drops \cite{Mai-new}. 
But they will also be helpful in our context.

\begin{figure*}[t!]
\includegraphics[width=18cm]{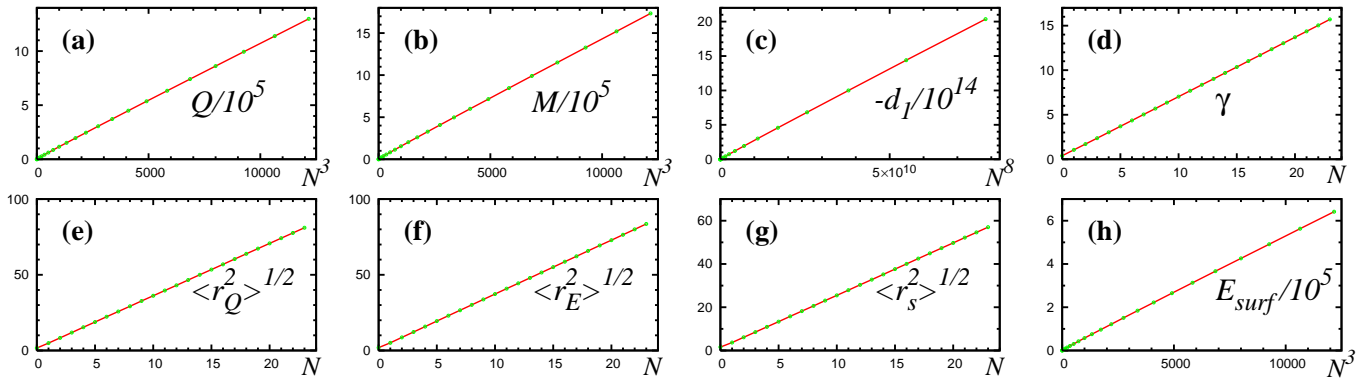}

\caption{\label{Fig-08:bulk-properties}
Various $Q$-ball properties plotted vs.\ $N^k$ with the power $k$
chosen according to the predictions in 
Eqs.~(\ref{Eq:N-behave-Q}--\ref{Eq:N-behave-ri2}).
The shown properties $X$ and the corresponding powers $k$,
written as pairs $(X,k)$ are:
(a) charge $(Q,3)$, 
(b) mass $(M,3)$, 
(c) $D$-term $(d_1,3)$, 
(d) surface tension $(\gamma,1)$,
the square roots of the mean square radii of the 
(e) charge distribution $(\la r_Q^2\ra^{1/2},1)$,
(f) energy distribution $(\la r_Q^2\ra^{1/2},1)$, 
(g) shear force distribution $(\la r_s^2\ra^{1/2},1)$, and 
(h) the surface energy $(E_{\rm surf},3)$. 
The discrete data sets are connected by lines to guide the eye.}
\end{figure*}

On the basis of the assumptions (ii, iii) we expect the charge $Q$ 
and mass $M$ to be proportional to the ``volume'' which grows like $N^3$
(even though the solutions are too diffuse to make ``volume'' a 
well-defined concept).
The scaling predictions (\ref{Eq:N-behave-ri2}) for the mean
square radii also follow straight forwardly from 
assumption (ii).

The prediction  (\ref{Eq:N-behave-Esurf}) for the surface energy
is at first glance counter-intuitive. One would expect $E_{\rm surf}$ to 
grow with ``surface area'' $\propto (\mbox{volume})^{2/3}\propto N^2$. 
However, we have to take into account the shell structure in point (i). 
The ground state has one surface, and the $N^{\rm th}$ excitation 
with its $N$ shells has in addition to that $2N$ surfaces. 
The contributions of individual surfaces do grow like $N^2$ 
as the size of the system grows $\propto N$ according to point (i).
But also the number of surfaces grows  $\propto N$,
which yields (\ref{Eq:N-behave-Esurf}). 
Similarly, we expect the surface tension 
$\gamma$ as defined in \cite{Coleman:1985ki,Mai-new} to be also 
proportional to the number of surfaces, hence the prediction
(\ref{Eq:N-behave-gamma}).

In order to derive the scaling behaviour of $d_1$ we may 
use dimensional arguments. The dimensionality of $d_1$ is 
$(\mbox{mass}\times\mbox{size})^2$ and with mass $\propto N^3$ 
and size $\propto N$  we obtain the prediction (\ref{Eq:N-behave-d1}).
Alternatively we may explore the liquid drop limit in which
$d^{\rm drop}_1=-\,\frac{4\pi}{3}\,M\,\gamma\,R^4$ where $R$ denotes 
the radius of the drop \cite{Polyakov:2002yz,Mai-new}. With the 
scaling predictions 
(\ref{Eq:N-behave-M},~\ref{Eq:N-behave-gamma},~\ref{Eq:N-behave-ri2})
for $M$, $\gamma$, and size of the system 
we are again lead to  the prediction (\ref{Eq:N-behave-d1}).

Fig.~\ref{Fig-08:bulk-properties} shows the global properties 
$Q$, $M$, $d_1$, $\gamma$, $\la r_i^2\ra$ for $i=Q,\,M,\,s$
and $E_{\rm surf}$ plotted as functions of $N^k$ with the powers $k$ 
as predicted in Eqs.~(\ref{Eq:N-behave-Q}--\ref{Eq:N-behave-ri2}).
The results fully confirm the predictions
(\ref{Eq:N-behave-Q}--\ref{Eq:N-behave-ri2}).
Hardly visible in Fig.~\ref{Fig-08:bulk-properties} is that for 
$N=0,\,1,\,2$ the global properties exhibit deviations from the 
scaling behavior (\ref{Eq:N-behave-Q}--\ref{Eq:N-behave-ri2}).
But for $N\gtrsim 2$ the numerical results follow 
Eqs.~(\ref{Eq:N-behave-Q}--\ref{Eq:N-behave-ri2}) with
very good accuracy, see Fig.~\ref{Fig-08:bulk-properties}.

\begin{figure}[b!]
\includegraphics[width=4.25cm]{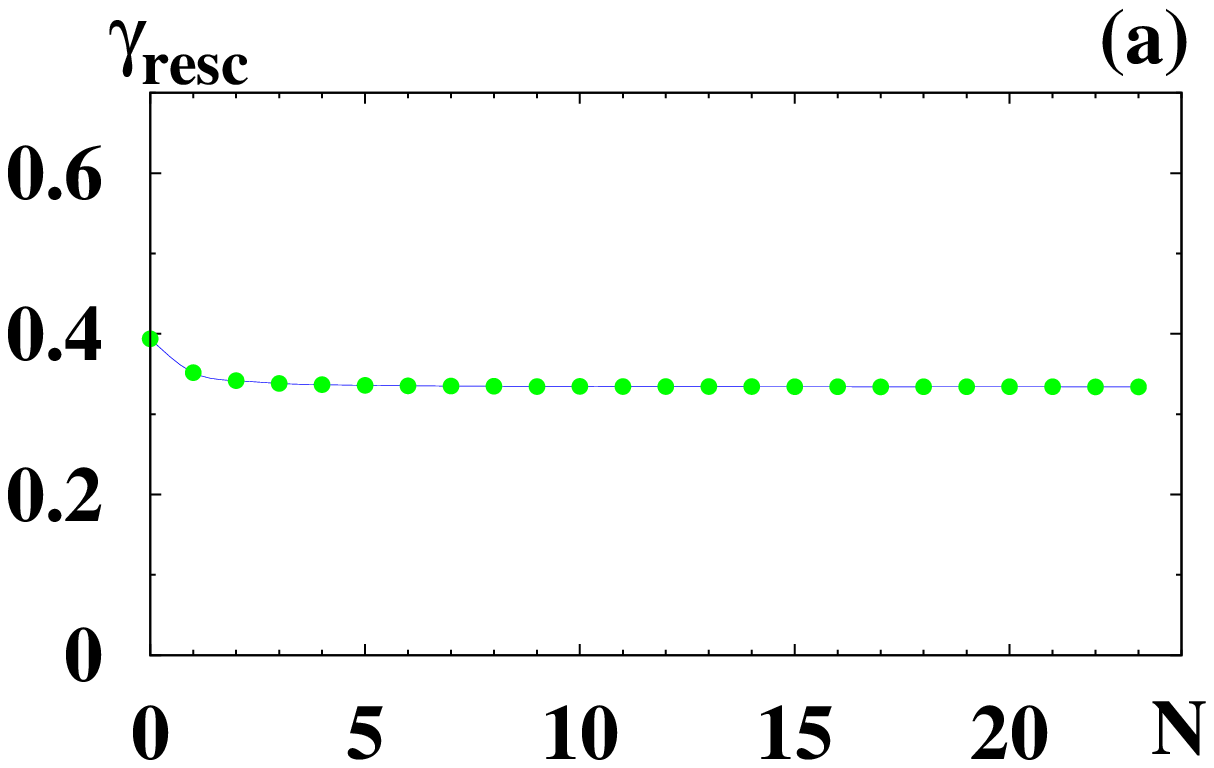}
\includegraphics[width=4.25cm]{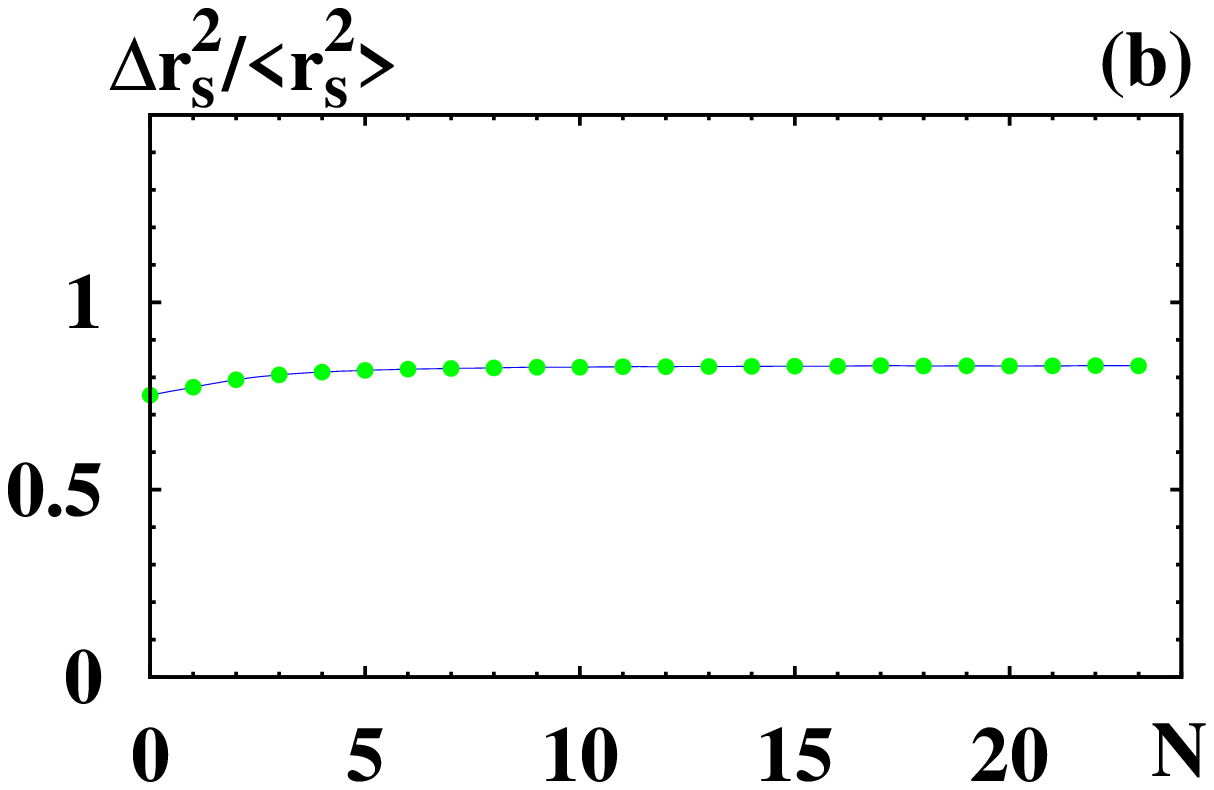}

\vspace{-3mm}

\caption{\label{Fig-09:diffuseness}
   (a) The true ``surface density'' $\gamma_{\rm resc}=\gamma/(2N+1)$
   as function of $N$. (b) The measure of the diffuseness of the system 
   $\Delta r_s^2/\la r_s^2\ra$ as function of $N$.}
\end{figure}

In particular, we observe $\gamma\propto N$ as predicted in
(\ref{Eq:N-behave-gamma}).
The more adequate property characterizing the ``surface tension'' 
at the ``boundary'' between $Q$-matter and vacuum is the rescaled 
quantity $\gamma_{\rm resc}=\gamma/(2N+1)$ which takes into account 
that the $N^{\rm th}$ radial excitation has $(2N+1)$ surfaces.
Fig.~\ref{Fig-09:diffuseness}a shows that $\gamma_{\rm resc}$
is nearly independent of $N$ as expected. 
We stress that  $\gamma_{\rm resc}$ is an average. $Q$-matter in 
excited $Q$-balls does not possess the same ``surface tension'' 
everywhere, otherwise the peaks in $s(r)$ in Fig.~\ref{Fig-05:shear} 
would be all equally high.

We would like to stress that although qualitatively 
here the liquid drop picture is useful, the concept 
of a surface tension is well justified only in the limit 
$\omega\to\omega_{\rm min}$ where the solutions exhibit
``sharp edges'' \cite{Coleman:1985ki}. For ground states
$\Delta r_s^2/\la r_s^2\ra$ can be used as a measure for 
the diffuseness of the system, where 
$(\Delta r_s^2)^2=\la r_s^4\ra-\la r_s^2\ra^2$ with
$\la r_s^4\ra=\int_0^\infty\di r\;r^4s(r)/\gamma$  \cite{Mai-new}. 
If for ground states $\Delta r_s^2/\la r_s^2\ra\ll 1$ 
one has ``sharp edges'' \cite{Mai-new}.
For our parameters $\Delta r_s^2/\la r_s^2\ra\simeq 0.75$ for $N=0$,
i.e.\ this condition is not convincingly realized; the system
is diffuse.
If we apply this measure also to excitations, we find that 
they are similarly diffuse to the ground state, 
see  Fig.~\ref{Fig-09:diffuseness}b.

For $\omega\to\omega_{\rm min}$ the $s(r)$ would 
become proportional to the sum of $(2N+1)$
$\delta$-functions with support at the positions of the surfaces 
of the shells
\cite{Mai-new}. If we assume for simplicity the surfaces equidistant and 
the coefficients of $\delta$-functions equal
(this is not accurate, see Sec.~\ref{Sec-3:densities}, 
but will be irrelevant after we take the limit 
$N\to\infty$ below) we would expect that 
$\la r_s^4\ra \propto \sum_{k=1}^{2N+1}k^4/(2N+1)$
while  $\la r_s^2\ra \propto \sum_{k=1}^{2N+1}k^2/(2N+1)$ and
\be\label{Eq:diffuseness-limit}
     \lim\limits_{N\to\infty}\,
     \frac{\Delta r_s^2}{\la r_s^2\ra}
     = \frac2{\sqrt{5}} \;\;\;\;\mbox{for}\;\;
     \omega\to\omega_{\rm min}.
\ee
This corresponds numerically to $0.894\dots$ 
and is remarkably close to the values observed for
$N\gtrsim 2$ in Fig.~\ref{Fig-09:diffuseness}b,
even though our $\omega$ is not close to $\omega_{\rm min}$.
It would be very interesting to test the prediction
(\ref{Eq:diffuseness-limit}) for $\omega$ closer to 
$\omega_{\rm min}$. But in such situations radial excitations are 
difficult to find numerically, see App.~\ref{App:technical-details}.

The shell structure can also be studied by looking at the charge 
distribution. Since the $\rho_{\rm ch}(r)$ vanish at the positions 
where the fields $\phi(r)$ change sign, this allows one to define 
exactly where a shell starts and where it ends. The last shell, 
of course, has no sharp boundary but vanishes exponentially 
according to (\ref{Eq:asymp-large}). 
Let us describe briefly how the charge is distributed in the largest 
excitation $N=23$ our numerical method could handle.
The inner region carries about $16.7\,\%$ of the total charge 
of this solution, the first shell $1.64\,\%$, and the second 
$1.58\,\%$ which is a global minimum. From here on the percentages 
carried by the subsequent shells increase gradually until the last 
shell contains $8.8\,\%$ of the total charge. 

\begin{figure}[b!]
\includegraphics[width=4.25cm]{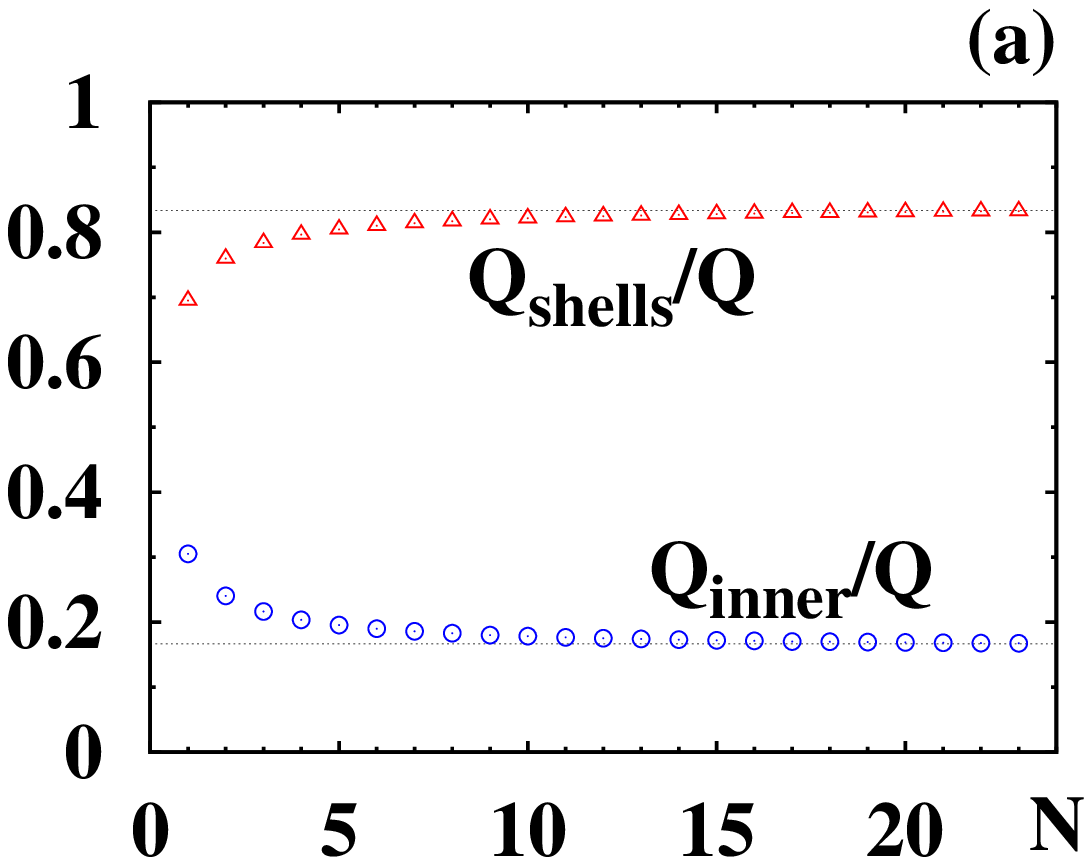}
\includegraphics[width=4.25cm]{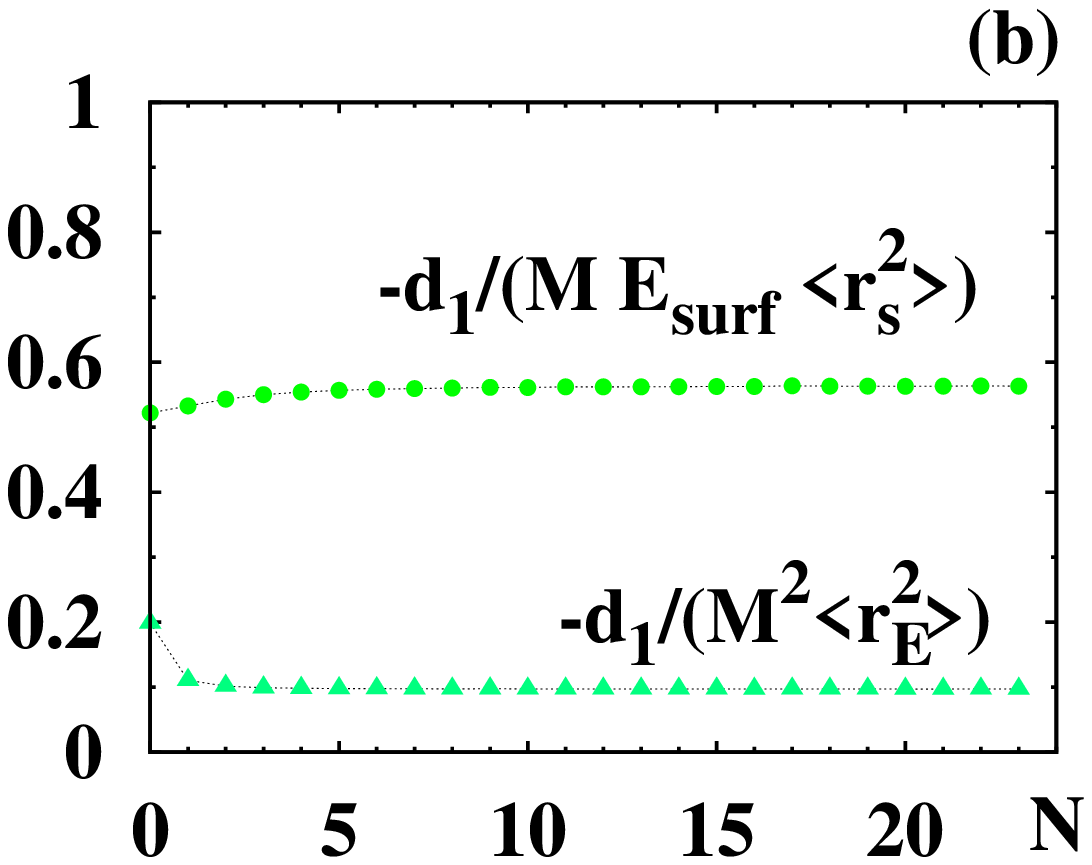}
\vspace{-3mm}

\caption{\label{Fig-10:Q-partition-d1-rescaled}
   (a)
   The relative contributions of the inner region (circles) 
   and the shell region (triangles) to the total charge as function
   of $N$.
   (b)
   The constant $d_1$ in units of $M^2\la r_E^2\ra$ (squares) and
   $M\,E_{\rm surf}\la r_s^2\ra$ (circles) as function of $N$.} 
\end{figure}

We did not observe regularities other that with respect to individual 
shells, but we found an interesting pattern how the charge is partitioned
between the inner region, and the shell region. Let us define 
$Q_{\rm inner}$ as the charge contained between $0\le r\le R_1$
where $R_1$ denotes the first zero of $\phi(r)$, and let 
$Q_{\rm shells}$ denote the charge carried by all shells,
such that $Q=Q_{\rm inner}+Q_{\rm shells}$.
The interesting observation is that
as $N$ increases $Q_{\rm inner}/Q\to\,\frac15$ from above,
while $Q_{\rm shells}/Q\to\,\frac45$ from below,
see Fig.~\ref{Fig-10:Q-partition-d1-rescaled}a.

Of all global properties studied in this work $d_1$ shows the strongest 
variations with $N$, as it did for ground states when $\omega$ was
varied \cite{Mai-new}. However, when taking the dimensionality of
$d_1$ into account, see above, one finds that the appropriately
scaled constant $d_1$ is bound from above and below.
In \cite{Mai-new} the following inequality was derived for all
solutions of the $Q$-ball equations of motion
\be\label{Eq:d1-inequality}
    0 < - \,\frac{d_1}{M^2\la r_E^2\ra} < \frac59 \;.
\ee
In Fig.~\ref{Fig-10:Q-partition-d1-rescaled}b we see that the radial 
excitations satisfy the inequality (\ref{Eq:d1-inequality}).

Finally let us mention the interesting relation of $d_1$ to the
relative ``wall width'' $\Delta r_s^2/\la r_s^2\ra$ derived in 
\cite{Mai-new} which can be expressed as 
\be\label{Eq:d1-relation-with-wall-width+N-limit}
     -\;\frac{d_1}{M\,E_{\rm surf}\la r_s^2\ra} = \frac13
     \Biggl(1+\biggl(\frac{\Delta r_s^2}{\la r_s^2\ra}\biggr)^{\!2}\Biggr)
     \,\;\;
     \stackrel{{\mbox{\footnotesize $N\to\infty\:\;\;\;\;$}\atop{
                \mbox{\footnotesize $\omega\to\omega_{\rm min}$}}}}
               {\mbox{\huge$\longrightarrow$}}
     \;\; \frac{3}{5}
     \;,\ee
where in the last step we used (\ref{Eq:diffuseness-limit}). Again, 
although in our calculation $\omega$ is not close to $\omega_{\rm min}$
we observe in Fig.~\ref{Fig-10:Q-partition-d1-rescaled}b 
that the numerical results are close to the limit derived in 
(\ref{Eq:d1-relation-with-wall-width+N-limit}).

\section{\boldmath Stability and $d_1$}
\label{Sec-V:stability-and-d1}

For all solutions we find $M<m\,Q$ where $m=\omega_{\rm max}$ denotes 
the mass of a $Q$-quantum. For the ground state this inequality implies 
absolute stability. But the radial excitations can decay.
For all our excitations lighter ground state configurations exist 
with the same charge. 

For example, our first excited state of 
$\omega^2=1.37$ has $Q=342$ and $M=461$. The following absolutely 
stable ground state solutions have the same total charge but a 
smaller total mass:
\begin{itemize}
\item one   $Q$-ball  of $\omega^2=0.51$ is  $1.61$ times lighter,
\item two   $Q$-balls of $\omega^2=0.61$ are $1.45$ times lighter, 
\item three $Q$-balls of $\omega^2=0.68$ are $1.35$ times lighter, 
$$ \vdots $$
\item fifteen $Q$-balls of $\omega^2=1.18$ are $1.008$ times lighter.
\end{itemize}
\noindent
The latter is the threshold for symmetric configurations, 
and 16 $Q$-balls with $\omega^2=1.21$ would be $0.5\,\%$ heavier.
Also asymmetric configurations with lower energy exist. 
E.g., the ground states for $\omega^2=0.516$ and $\omega^2=1.37$ (i.e.\ the
groundstate of our excitation) have the same total charge but are 1.55
times lighter than the first excited state of $\omega^2=1.37$.
The still heavier excitations $N>1$ have accordingly more 
decay modes.\footnote{
  Here we content ourselves to observe that more stable 
  configurations exist, and are not concerned with the
  dynamics of the possible decays. All numbers quoted
  for $\omega^2\neq1.37$ are from \cite{Mai-new}.}
In short, all radial excitations are unstable.

\begin{figure*}[t!]
\includegraphics[width=18cm]{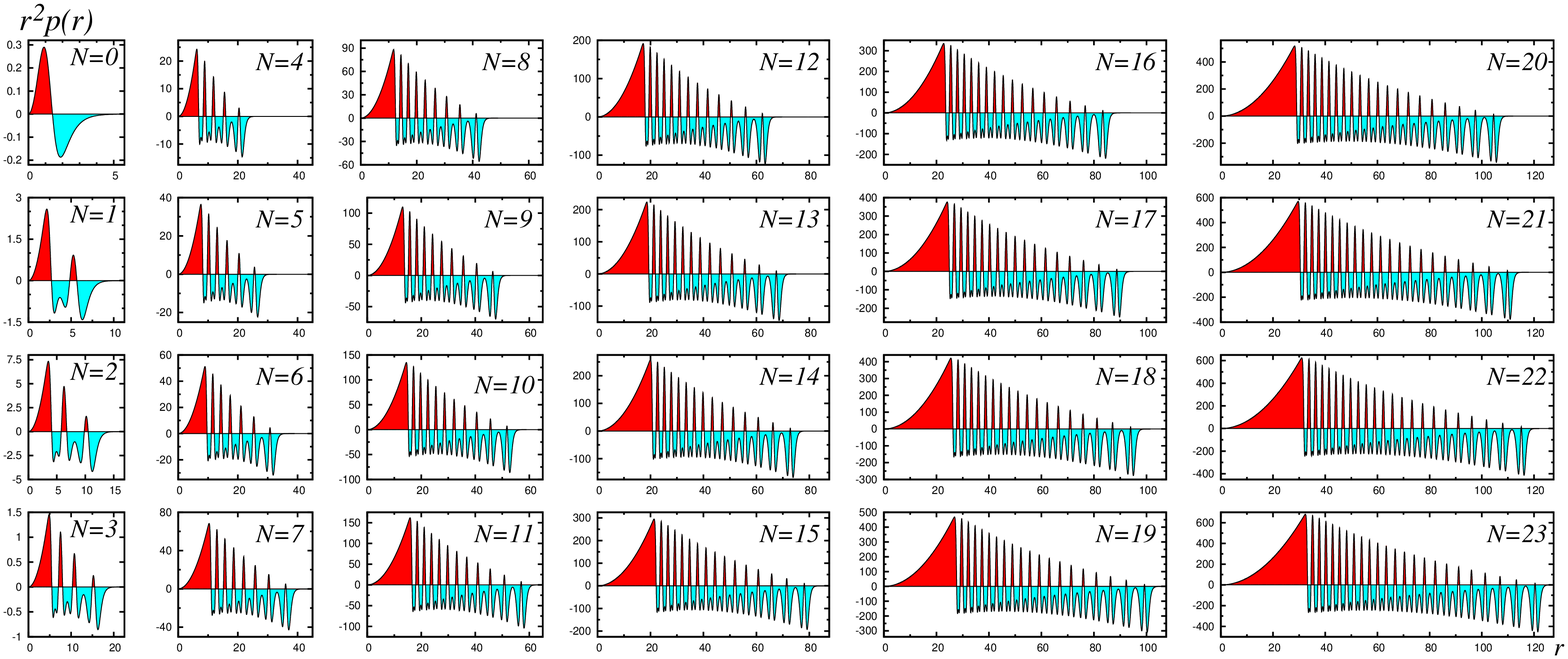}

\vspace{-3mm}

\caption{\label{Fig-12:p-r2}
  $r^2p(r)$ as functions of $r$ for $0\le N \le 23$.
  Except for the first column the scales on the $r$-axis are
  kept constant for a better comparison. The shaded regions
  above and below the $r$-axis have equal areas such that
  $\int_0^\infty\di r\,r^2p(r)=0$.}

\vspace{8mm}

\includegraphics[width=18cm]{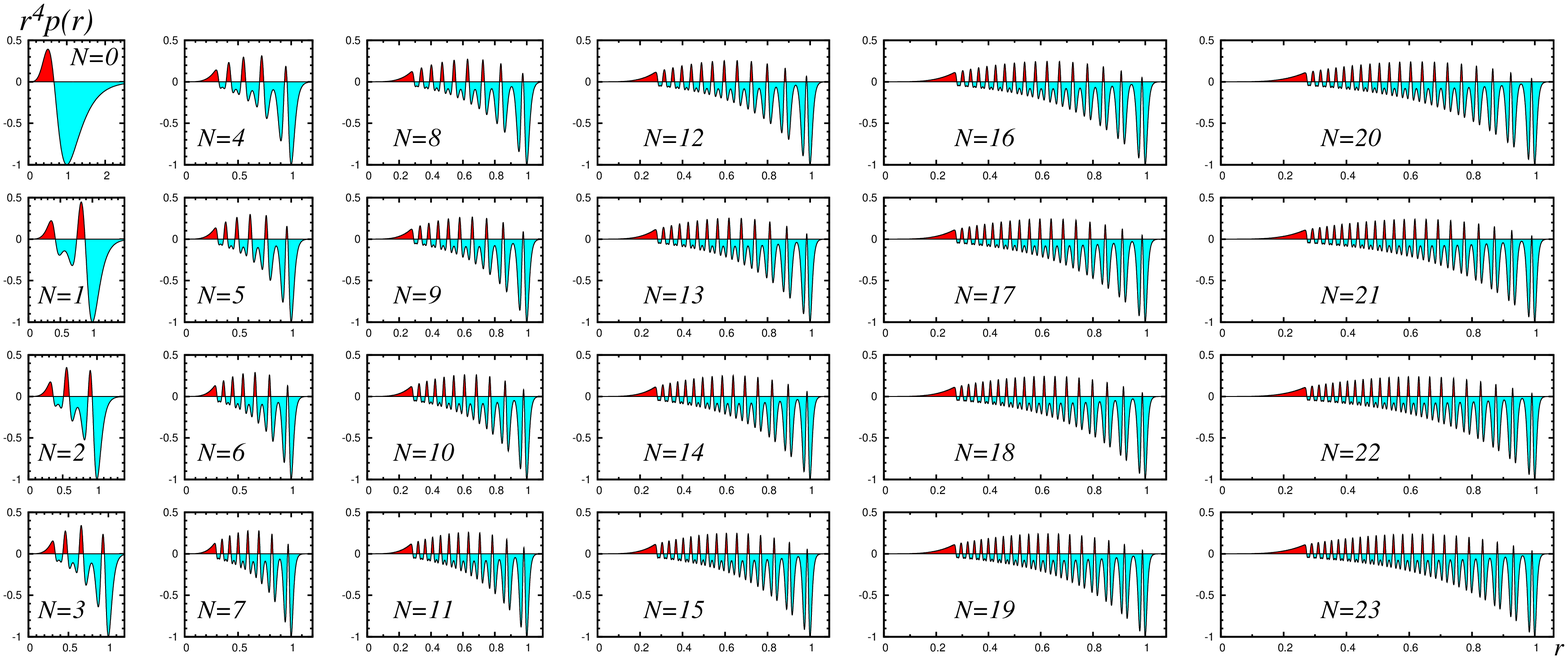}
\caption{\label{Fig-13:p-r4}
   $r^4p(r)/C_{\rm last}$ as functions of $r/R_{\rm last}$ 
   for $0\le N \le 23$, where $R_{\rm last}$ denotes the position
   of the last minimum of $r^4p(r)$ and 
   $C_{\rm last}=|R_{\rm last}^4p(R_{\rm last})|$.
   With these units the global of all curves occur at 
   $r/R_{\rm last}=1$ and assume the value $r^4p(r)/C_{\rm last}=-1$ 
   which makes a comparison easier. $r^4p(r)$ is the 
   integrand of $d_1$. The figure demonstrates how 
   the negative sign of $d_1$ appears.}

\end{figure*}

Nevertheless, the solutions with $N>0$ of course also minimize the 
energy functional, though they correspond to local minima of 
the action. One way to test this offers the stability condition,
or ``von Laue--condition'' \cite{von-Laue},
\be\label{Eq:stability-condition}
      \int\limits_0^\infty \di r\;r^2p(r)=0\;,
\ee
which was proven to be satisfied for all finite energy
solutions in the $Q$-ball system in \cite{Mai-new}.
It furthermore was shown that for all finite energy solutions
the pressure is positive for small $r$ and negative for
large $r$ \cite{Mai-new}. 
In Sec.~\ref{Sec-3:densities} we have seen that the pressure
distribution of the $N^{\rm th}$ radial excitation exhibits
this pattern and changes sign $(2N+1)$ times.
It is instructive to look in some more detail how excited 
$Q$-balls realize the condition (\ref{Eq:stability-condition}).

Fig.~\ref{Fig-12:p-r2} shows $r^2p(r)$ as function of $r$ for
the ground state and the radial excitations. In spite of the
complexity of the results the condition (\ref{Eq:stability-condition})
is satisfied within numerical accuracy which can be quantified as
follows. For instance, for the ground state we obtain
$|\int_0^\infty\di r\,r^2p(r)|/\int_0^\infty\di r\,r^2|p(r)|={\cal O}(10^{-8})$
and similarly up to $N\le 4$. 
With increasing $N$ it becomes more difficult to maintain this
accuracy. For $N\gtrsim 10$ the accuracy is 
in the range ${\cal O}(10^{-5})$ to ${\cal O}(10^{-3})$.

The regions with positive pressure provide forces directed
towards outside. These repulsive forces are compensated by 
negative pressure regions with attractive
forces directed towards the center. Repulsive
and attractive forces cancel precisely according
to (\ref{Eq:stability-condition}).

It is interesting to note that the role of the shells is
to compensate the repulsive forces from the core. In fact,
on average the shells contribute attractive forces. 

In Ref.~\cite{Mai-new} it was shown that the pattern how 
the pressure distribution satisfies the condition 
(\ref{Eq:stability-condition}) at once implies that the
constant $d_1$ must have a negative sign. Although the
sign of $d_1$ can also be deduced from the shear forces
\cite{Mai-new}, this indicates a connection between $d_1$
and stability. In \cite{Mai-new} only $Q$-ball ground 
states were studied, for which $p(r)$ changes sign only once.
Nevertheless the general proof that the stability  condition 
(\ref{Eq:stability-condition}) implies $d_1<0$ in \cite{Mai-new} 
was formulated assuming that the pressure change
the sign an arbitrary odd number of times. 
This is the situation we encounter for radial excitations,
and our results illustrate how the stability  condition 
(\ref{Eq:stability-condition}) determines the sign of $d_1$.

Fig.~\ref{Fig-13:p-r4} shows $r^4p(r)$ as functions of $r$.
Clearly, integrating this function over $r$ yields a 
negative number, and up to a prefactor of $5\pi M$ 
the constant $d_1$, cf.\ Eq.~(\ref{Eq:def-d1}).
Our results for the pressure distribution therefore
fully confirm the general proof of the negative sign of $d_1$ 
from the stability relation (\ref{Eq:stability-condition})
formulated in \cite{Mai-new}.

We remark that in the proof of \cite{Mai-new}
also the possibility was considered that the $p(r)$ could
become zero at some point without changing sign.
We do not encounter this situation for the parameters
used in this work.

\section{Conclusions}
\label{Sec-6:conclusions}

We presented a study of the energy momentum tensor 
of $Q$-balls in a scalar field theory with U(1) symmetry.
While in a previous work we investigated in detail ground 
state solutions for different $\omega$ \cite{Mai-new}, in 
this work radial excitations of $Q$-balls were in the focus
of our study.

In Ref.~\cite{Volkov:2002aj} the radial excitations 
$N=1, \,2$ were studied previously for fixed charge $Q$,
in other words the excitations were classified by specifying
the charge and order $(Q,N)$.
Here we adopted a different classification scheme and fixed 
$\omega$, i.e.\ the excitations are specified by $(\omega,N)$.
We were able to find numerically solutions for the ground state 
$N=0$ and $1\le N \le 23$ excitations.
All solutions obtained in this work were exact solutions of the 
equations of motion. The numerical results were subject to
stringent tests to guarantee their correctness. 
On the basis of our results reaching high in the spectrum 
of radial excitations we were able to obtain fascinating 
insights in the structure of these excitations.

As $N$ grows the systems exhibit increasing degrees of
complexity. The radial field of the $N^{\rm th}$ excitation 
changes the sign $N$-times. At the positions $R_i$ with
$1 \le i \le N$ where this happens the otherwise positive
charge distribution vanishes exactly, and the energy
density shows at positions very close to the $R_i$ 
clear minima. In other words, the charge is distributed over
an inner region with nearly constant density surrounded
by $N$ shells, and $T_{00}(r)$ closely follows this pattern.
We observed the interesting pattern that, as $N$ increases, 
the constant density inner region carries $1/6$ of the
total charge, while the remaining $5/6$ are distributed
on the shells.

The energy densities show in addition also characteristic
spikes at the ``edges'' of the shells due to the impact of the 
``surface energy.'' The effects of the ``surface tension'' are 
reflected with even more clarity in the shear force distributions. 
We have shown that the system is diffuse for the parameters considered 
in this work, and discussed in which sense the concepts 
``surface tension'' and ``surface energy'' are nevertheless useful. 
The highest degree of complexity is seen in the pressure distributions
which change the sign $(2N+1)$ times. 

\

In spite of the complexity of the solutions, the properties of the
excited $Q$-balls scale with $N$ with great regularity. 
For instance, the size of the system is proportional to $N$,
independently whether one uses the zeros of the $\phi(r)$
or square roots of various mean square radii to define it.
On the basis of general arguments we were able to predict
also the scaling of other quantities, for instance
$M\propto N^3$ or $d_1\propto N^8$ which are supported
by our numerical results.
Remarkably, among all quantities we studied, the $D$-term 
varies most strongly with $N$. Similarly $d_1$ was the 
quantity which varied most strongly in the study of ground 
state solutions as functions of $\omega$ \cite{Mai-new}.

One of the consequences of EMT conservation is the stability 
(or von-Laue-) condition \cite{Polyakov:2002yz,von-Laue} 
stating that  $\int_0^\infty \di r\:r^2p(r)=0$. 
In \cite{Mai-new} this condition was proven analytically to be 
satisfied for any solution of $Q$-ball equations of motion,
and in this work we could verify numerically that the 
$p(r)$ of radial excitations with its $(2N+1)$ precisely
integrates to zero with very good numerical precision.

The important result is that the $D$-term is negative also
for all radial excitations. In all approaches where $d_1$
was studied so far, it was found negative.
But only the $D$-terms of ground states were studied so far,
and to best of our knowledge this is the first time 
excited states are shown to have also negative $D$-terms.

In  \cite{Mai-new} a rigorous proof was given that for 
$Q$-balls $d_1<0$ follows from the stability condition 
and $Q$-ball equations of motion. In  \cite{Mai-new} only ground 
state solutions were studied for which $p(r)$ changes sign only once. 
Nevertheless the proof had to be formulated assuming that $p(r)$ 
could more generally change the sign any odd number of times.
The results obtained in this work illustrate that this is
not a pathological case which has to be taken into account 
for the sake of mathematical rigor. Indeed, for excited 
$Q$-balls one does encounter such a situation in practice. 

In this work we also fully confirm another finding of 
\cite{Mai-new}, namely that stability is a sufficient 
but not necessary condition for $d_1$ to be negative.
In fact, we have shown that all radial excitations
obtained in this work are unstable. They correspond to
local but not global minima of the action, and can decay 
into configurations of absolutely stable ground states 
with the same total charge but a smaller total mass.

The works presented here and in \cite{Mai-new} clearly 
demonstrate the property $d_1<0$ for $Q$-ball systems
and, we hope, will inspire rigorous proofs of this
property also in other systems. Our results also establish 
$d_1$ as a particle property particularly sensitive to 
variations of parameters of the system. An interesting
question remains: can $d_1$ be ever positive in a
physical system? 

{}

\

\noindent{\bf Acknowledgements.}
We thank Gerald Dunne and Alex Kovner for helpful discussions.
The work was partly supported by DOE contract DE-AC05-06OR23177, under
which Jefferson Science Associates, LLC,  operates the Jefferson Lab.

\newpage 

\appendix
\section{Technical details}
\label{App:technical-details}

We assume $\omega > 0$ without loss of generality. Finite energy 
solutions exist for $\omega$ in the range \cite{Coleman:1985ki}
\be\label{App-Eq:condition-for-existence}
      \omega_{\rm min}^2 \equiv
      \min\limits_\phi\biggl[\frac{2\,V(\phi)}{\phi^2}\biggr]
      < \omega^2 < \omega_{\rm max}^2 \equiv 
      V^{\prime\prime}(\phi)\biggl|_{\phi=0}\;.\ee
For the potential used in this work $0.2 < \omega^2 < 2.2$.
The ground states are absolutely stable 
for $\omega^2<\omega^2_{\rm abs}\approx 1.55$ \cite{Mai-new}.
For $\omega$ close to $\omega_{\rm min}$ it is numerically 
challenging to handle the ground states, let alone radial 
excitations. In order to have an absolutely stable ground state, 
and maximize the chances find numerous radial excitations it is 
profitable to work close to $\omega^2_{\rm abs}\approx 1.55$.
In this sense, $\omega=\sqrt{1.37}\approx 0.94\,\omega_{\rm abs}$ 
is among the ideal choices.
%

As $N$ increases, see Sec.~\ref{Sec-2:radial-excitations}, it is necessary
to release the particles close to the maximum of $U_{\rm eff}$
given by 
\be\label{App-Eq:constant-solution}
      \phi_{\rm const}(\omega) = \sqrt{\frac{B}{C}\,\biggl(\frac13+\frac16\,
      \sqrt{1+\frac{6C}{B^2}(\omega^2-\omega_{\rm min}^2)}\,\biggr)}\;,
\ee
where the subscript reminds that (\ref{App-Eq:constant-solution})
corresponds to one of the ``stationary'' solutions $\phi(r)={\rm const}$
of (\ref{Eq:eom}) \cite{Mai-new}, which however do not satisfy the 
boundary condition for $r\to\infty$. For our parameters 
$\phi_{\rm const}=1.1045\dots$ and the radial excitations
$N\ge 1$ are all within $10^{-6}$ of this value.

\section{Numerical tests}
\label{Sec:test-numerics}

In view of the complexity of the solutions, it is important to
monitor the numerical quality of the solutions. For that we
made the following tests. We checked that

\begin{itemize} 

   \item[A.]   the stability condition (\ref{Eq:stability-condition}) is valid,

   \item[B.]   the equation
           $\frac{2}{r}\,s(r) + \frac{2}{3}\,s^\prime(r) + p^\prime(r) = 0$
           is satisfied,

   \item[C.]  the expressions for $d_1$ in (\ref{Eq:def-d1})
              yield the same result, 

   \item[D.]   $p(0) = 2\int_0^\infty\di r\;\frac{s(r)}{r}$ is equal
               to $p(0)$ from (\ref{Eq:pressure-density}).

\end{itemize}
All these relations can be derived from EMT conservation 
\cite{Polyakov:2002yz,Goeke:2007fp} and provide powerful tests
for the numerics \cite{Mai-new}. 
We find relative numerical accuracies between ${\cal O}(10^{-9})$ 
and ${\cal O}(10^{-3})$ depending on $N$ and the kind of test.

In Sec.~\ref{Sec-V:stability-and-d1} we already reported  
how the stability condition, test (A), is satisfied numerically.
For (B) we checked that 
$(\frac{2}{r}\,s(r) + \frac{2}{3}\,s^\prime(r) + p^\prime(r))/
 (\frac{2}{r}\,|s(r)| + \frac{2}{3}\,|s^\prime(r)| + |p^\prime(r)|)$
is typically of ${\cal O}(10^{-3})$ or smaller, for $r>0$ and $\forall\, N$. 

Concerning test (C): for instance, for the highest excitation 
$N=23$ we were able to handle with our numerics,
we obtain from (\ref{Eq:def-d1}):
$d_1^p=-2.0366\times 10^{15}$ using pressure distribution vs.\
$d_1^s=-2.0360\times 10^{15}$ from shear forces,
which corresponds to a relative accuracy of $3 \times 10^{-4}$.

Concerning test (D): we obtain e.g.\ for $N=23$ the result
$p(0)=0.654652$ from Eq.~(\ref{Eq:pressure-density}),
while using the above quoted formula yields $p(0)=0.654655$, 
which corresponds to a relative accuracy of $5\times10^{-5}$. 

On the basis of these stringent tests we are confident 
that none of the bumps, peaks, structures in 
Figs.~\ref{Fig-02:phi}--\ref{Fig-13:p-r4}
are numerical artifacts, but all details of our numerical 
solutions reflect the true characteristics of the excited states.


\end{document}